\documentclass[12]{article}
\usepackage[utf8]{inputenc}
\usepackage[T1]{fontenc}
\usepackage{graphicx}
\usepackage{authblk}
\usepackage{siunitx}
\usepackage{helvet}
\usepackage{amsmath}

\usepackage{geometry}
\usepackage{textgreek}
\usepackage{nameref}
\usepackage{nicefrac}
\usepackage{comment}
\usepackage{siunitx}
\usepackage{hyperref}
\usepackage{xcolor}
 \usepackage{xcolor}

\usepackage[
backend = biber,
style=numeric,
sorting=none,
doi=false,isbn=false,url=false,
maxbibnames=5 
]
{biblatex}
\DeclareNameAlias{author}{family-given}
\DeclareFieldFormat[article,periodical]{volume}{\mkbibbold{#1}}

\usepackage{lineno}

\title{Numerical model for 32-bit magnonic ripple carry adder}

\author[1]{U. Garlando}
\author[2]{Q. Wang}
\author[2]{O. V. Dobrovolskiy}
\author[2]{A. V. Chumak}
\author[1]{F. Riente\thanks{Corresponding author.}}
\affil[1]{Department of Electronics and Telecommunications, Politecnico di Torino, Corso Duca degli Abruzzi 24, 10129 Torino, Italy}
\affil[2]{Faculty of Physics, University of Vienna,
Boltzmanngasse 5, A-1090 Vienna, Austria}

\date{}
\begin{document}

\maketitle

 
\begin{abstract}
\normalsize
In CMOS-based electronics, the most straightforward way to implement a summation operation is to use the ripple carry adder (RCA). Magnonics, the field of science concerned with data processing by spin-waves and their quanta magnons, recently proposed a magnonic half-adder that can be considered as the simplest magnonic integrated circuit. Here, we develop a computation model for the magnonic basic blocks to enable the design and simulation of magnonic gates and magnonic circuits of arbitrary complexity and demonstrate its functionality on the example of a 32-bit integrated RCA. It is shown that the RCA requires the utilization of additional regenerators based on magnonic directional couplers with embedded amplifiers to normalize the magnon signals in-between the half-adders. The benchmarking of large-scale magnonic integrated circuits is performed. The energy consumption of 30 nm-based magnonic 32-bit adder can be as low as 961 aJ per operation with taking into account all required amplifiers.
\end{abstract}
\vspace{1cm}

\section*{Introduction}

Over the last years, spin-waves (SWs) and their quanta -- magnons -- have attracted much attention due to their potential applications as data carriers in future data processing technologies \supercite{Barman2021,Pirro2021,Dieny2020,Chumak2015,Krawczyk2014,Wang2020}. Spin-waves are propagating disturbances in the spin order of a solid body which occurs without any motion of electrons and, thus, without Joule heating \supercite{Kajiwara2010,Collet2017,Heinz2020,Liu2018}. Moreover, the phase of spin-wave provides additional degrees of freedom (beyond amplitude) to code information, and the features of waves (de/constructive interference, diffraction, etc.) simplify the design structure of wave-based logic gates \supercite{Wang2020,Goto2019,Fischer2017,Talmelli2020}. Furthermore, the nanoscale wavelength and pronounced nonlinear phenomena of spin-waves are unique comparing to the acoustic waves and microwaves \supercite{Wintz2016,Che2020,Krivosik2010,Verba2019,} that makes them promising for the nanoscale Boolean/non-Boolean computing \supercite{Papp2021,Brächer2018,papp2020,Csaba2017}. 

Several magnonic devices have already been demonstrated at the early stage of single logic gate level, including spin-wave logic gates \supercite{Lee2008,Schneider2008}, majority gates \supercite{Fischer2017,Talmelli2020} and magnon transistors \supercite{Chumak2014,Wu2018}. In general, one can define two main approaches for the construction of magnonic circuits: the first one can be named a "converter-based" and relies on the utilization of highly efficient magnon-to-current converters used after each operation with data \supercite{Dutta2015,Egel2017,Mahmoud2020}. The magnonic circuits based on this approach are described in the section "Magnonic circuits" by C. Adelmann, et al. in \supercite{Barman2021}. The other approach is named "all-magnon" and, although some conversion from magnon to current is still required, aims for the minimization of the converters number via the utilization of natural strongly-pronounced magnonic nonlinear phenomena \supercite{Chumak2014}. Recently, a nanoscale magnonic directional coupler was realized, and its nonlinear functionality was demonstrated experimentally \supercite{Wang2020}. Furthermore, it was shown numerically that a magnonic half-adder, consisting of an XOR logic gate and AND logic gate, can be realized by combining two directional couplers into a circuit. The half-adder was specially designed to be applicable for further integration after a low-energy amplifier is added \supercite{Wang2020}. Nevertheless, the circuitry which would allow for synchronous operation of many such devices together to perform complex arithmetic operations was far beyond the scope of the previous investigations since such simulations are computationally too expensive and unfeasible.

Here, we present a numerical model which allows for the realization of complex all-magnon circuits based on the functional blocks of the previously studied half-adder \supercite{Wang2020}. The model is demonstrated on the example of a 32-bit integrated ripple carry addder. We conclude that complex magnon circuits require the utilization of additional regenerators with embedded amplifiers to restore degraded magnon signals in-between the half-adders. The benchmarking of large-scale magnonic integrated circuits is performed.

\section*{Results and Discussion}
\subsection*{Magnonic adder structure and operational principle}
\begin{figure}[ht]
\centering
  \includegraphics[width=\linewidth]{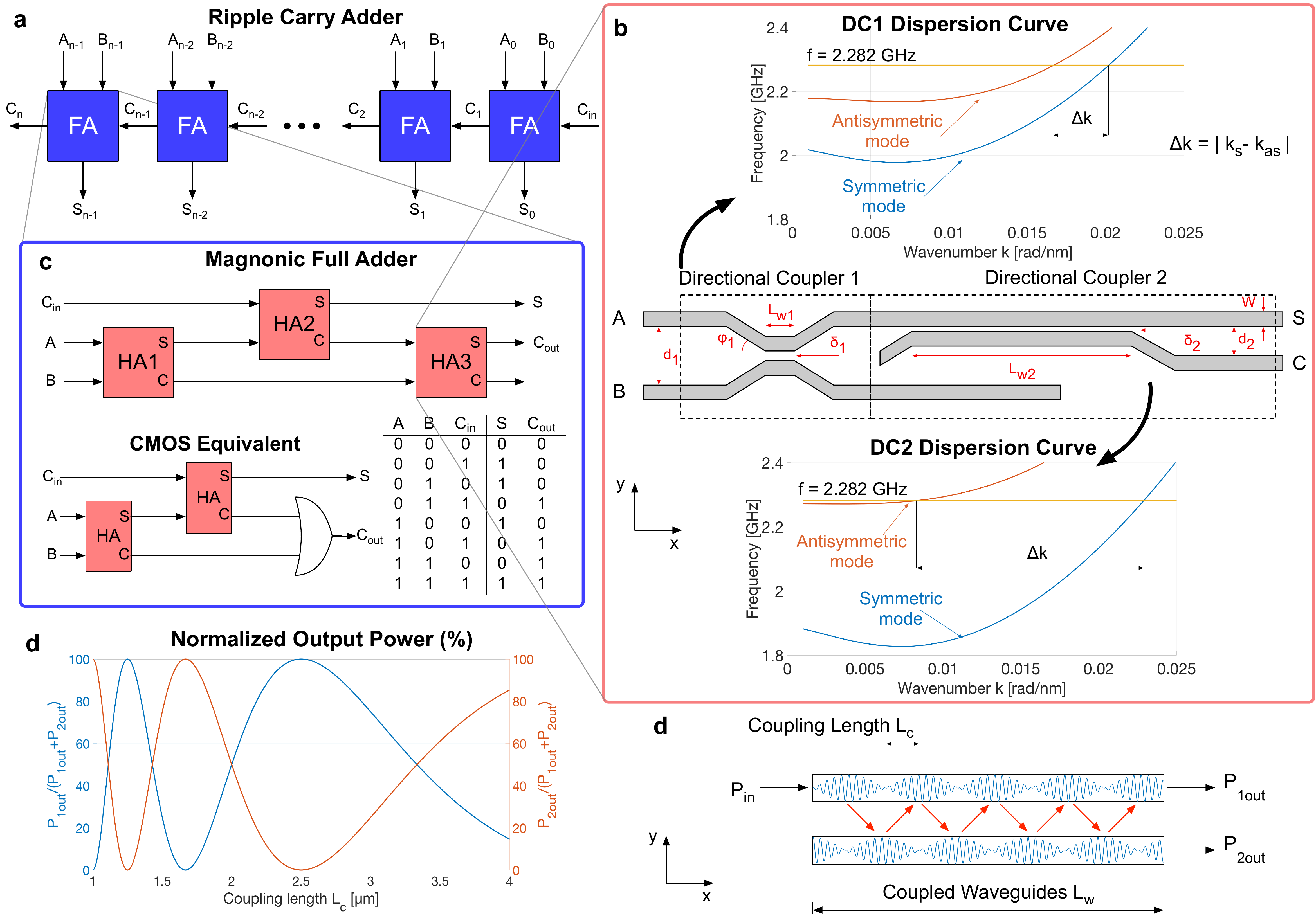}
  \caption{Sketch of the investigated system: \textbf{a}) N-bit ripple carry adder design; \textbf{b}) Zoom of the most important magnonic block, the half adder. The top and bottom graphs depict the dispersion curves of the "symmetric" (s) and "antisymmetric" (as) lowest collective spin-wave modes of a pair of coupled waveguides, the DC1 (operating in the linear regime) and the DC2 (operating in the non-linear regime), respectively. The central image depicts the design of the magnonic HA where all the dimensions involved are reported; \textbf{c}) Schematic representation of a magnonic full adder (top) and its CMOS equivalent representation (bottom) with the corresponding truth table; \textbf{d}) Normalized output power as a function of the coupling length $L_c$ for a fixed length of the coupled waveguides $L_w = 4\,\mu$m without damping; \textbf{d}) Schematic representation of the periodic energy exchange between two coupled spin-wave waveguides.}
  \label{fig:fig_1b} 
\end{figure}

Among combinational circuits, the most straightforward way to implement a summation is to use the ripple carry adder (Fig.~\ref{fig:fig_1b}.a). In many computer architectures, adders are used in the arithmetic logic unit and other processor parts. The fundamental element of such an adder is the full adder (FA). Multiple full adders can be cascaded in parallel to add N-bit operands. As suggested by its name, the carry-out bit is rippled into the next stage in this implementation. The full adder adds binary numbers, particularly it sums three inputs (A\textsubscript{i}, B\textsubscript{i}, C\textsubscript{i-1}) and produces two outputs (S\textsubscript{i}, C\textsubscript{i+1}), which represent the sum and the carry-out, respectively. It can be implemented in many ways, and one example is reported in Fig.~\ref{fig:fig_1b}.b. The structure depicted in Fig.~\ref{fig:fig_1b}.b is based on the half adder (HA), which is the most important magnonic building block to perform logic computation \supercite{Wang2020}.
In particular, the design proposed in this paper uses three magnonic half adders. The implemented logic function is reported in Eq.~(\ref{eq:fulladder_s}) and (\ref{eq:fulladder_c}), for the sum and carry respectively, where the over brackets represent the operation performed by every HA.
\begin{align}
\label{eq:fulladder_s}
    S &= \overbrace{\overbrace{A \oplus B}^\text{HA1} \oplus  C_{in}}^\text{HA2} \\
\label{eq:fulladder_c}
    C_{out} &= \overbrace{ \overbrace{(A \cdot B)}^\text{HA1} + \overbrace{(C_{in} \cdot \overbrace{(A \oplus B)}^\text{HA1})}^\text{HA2}}^\text{HA3}
\end{align}

From Fig.~\ref{fig:fig_1b}.c it is possible to observe that the third magnonic HA is used only as OR gate. For the sake of clarity, the CMOS equivalent and its truth table are reported in Fig.~\ref{fig:fig_1b}.c.

The magnonic half adder is composed of two directional couplers (DCs), one operating in the linear regime and the other in the non-linear regime\supercite{Wang2020}, they are named DC1 and DC2 respectively (Fig.~\ref{fig:fig_1b}.b). In both cases, the dispersion curve splits into symmetric (s) and antisymmetric modes (as) due to the dipolar interaction between the parallel waveguides. When the excited spin-wave is above the minimum of the antisymmetric mode ($f = 2.282\,$GHz in Fig.~\ref{fig:fig_1b}.b) both modes can be excited simultaneously in the coupled waveguides. The two modes have the same frequency but different wavenumber ($k_s$, $k_{as}$) that result in a different phase accumulation. The interference of these two modes result in energy exchange between the dipolar coupled waveguides. There is a periodic exchange of energy between the spin-waves in one waveguide to the other and vice-versa, which is named coupling length $L_{c}$. This phenomenon is schematically represented in Fig.~\ref{fig:fig_1b}.d and can be calculated as:
\begin{equation}
    L_c = \dfrac{\pi}{\Delta k_x} = \dfrac{\pi}{|k_s - k_{as}|}
\end{equation}
The coupling length depends on different parameters such as the spin-wave wavelength, spin-wave power, and geometrical parameter of the waveguide \supercite{Sadovnikov2016,Wang2018,Sadovnikov2017}.
The DC1, working in the linear regime, operates as a power splitter, while the DC2, working in the non-linear regime, operates both as AND/XOR gate\supercite{Wang2020}. Fig.~\ref{fig:fig_1b}.d shows the normalized output power in the coupled waveguides as a function of the coupling length $L_c$. It can be expressed using the Eq.~(\ref{eq:powersplit}).
\begin{equation}
\label{eq:powersplit}
    \dfrac{P_{1out}}{P_{1out} + P_{2out}} = \cos^2\bigg(\dfrac{\pi L_w}{2 L_c}\bigg),
\end{equation}
where $L_w$ represents the length of the coupled region. The normalized output power expression shows that the length of the coupled region and the coupling length play a crucial role in terms of power splitting and the functionality of the directional couplers.

\subsection*{Compact physical model}
\label{sec:model}
The most accurate approach to obtain the dispersion relation of two coupled waveguides is to solve the Landau-Lifshitz equation for the magnetization dynamics\supercite{Wang2018}. However, this approach is too complex and computationally expensive to be integrated within a tool for circuit-level exploration. On the other hand, the compact physical model presented keeps high accuracy, providing the flexibility to explore magnonic circuits considering their physical properties. It is openly available on Zenodo\supercite{riente_zenodo_2021}.

The model we developed describes the dispersion relation of the DC1 and the DC2 depending on the geometrical characteristics of the couplers and the spin-wave amplitude. It considers damping losses and the non-uniform width profile of the fundamental spin-wave mode of the waveguide\supercite{Wang2018,Guslienko2005,Verba2012}. The effective width ($w_{eff}$) of the waveguide can be larger than the nominal width ($w$) when the effective pinning decreases\supercite{Wang2018}. A change in the dispersion curve results in a variation of the coupling length $L_c$ and, as a consequence, in a different output power partition.
The expression for computing the dispersion relation of two coupled waveguides is reported in Eq.~(\ref{eq:dispersion})
\begin{equation}
\label{eq:dispersion}
f_{a,as}^{nl} (k_x, a_k) = f_{s,as}^0 (k_x) + T_k |a_k|^2,
\end{equation}
where $f_{s,as}^0 (k_x)$ represents the dispersion relation for the symmetric and antisymmetric spin-wave modes in coupled waveguides at linear region, $T_k$ is the nonlinear frequency shift coefficient \supercite{verba2016} in the isolated waveguide and $a_k$ is a dimensionless quantity and represents the spin-wave amplitude (see \nameref{sec:method}).

\begin{figure}[tbh]
\centering
  \includegraphics[width=0.6\linewidth]{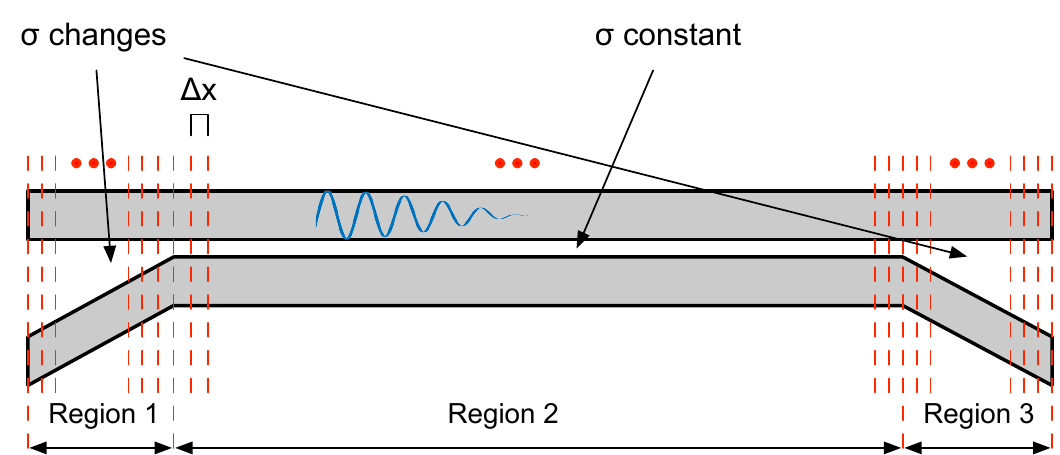}
  \caption{The directional coupler can be divided into three regions: regions 1 and 3 show an increasing/decreasing distance between the two waveguides and region 2 where the gap is constant. The main contribution to the dispersion curve comes from region 2. The introduced discretization along the $x$ axis makes it possible to take into account the additional coupling introduced by the opening/closing arms (region 1 and region 3).}
  \label{fig:fig2} 
\end{figure}

In general, the direction couplers are the core elements in magnonic circuits consisting of three main regions. Two oblique branches (opening/closing arms) are represented by regions 1 and 3 in Fig.~\ref{fig:fig2} and the region 2 shows the coupled region where the waveguides are parallel to each other. Most of the energy exchange between the coupled waveguides is observed in region 2, where the gap $\sigma$ is very small, 50\,nm and 10\,nm for the DC1 and DC2 respectively (100\,nm technology node). However, there is an additional contribution coming from regions 1 and 3. Starting from the power partition formula reported in Eq.~(\ref{eq:powersplit}) it is possible to define the number of jumps along the coupled region 2 as $N = L_w/L_c$.
Substituting in Eq.~(\ref{eq:powersplit}), the equation can be rewritten as $\cos^2 (\dfrac{\pi}{2} N)$. The number of jumps from one waveguide to the other is defined by the subsequent constructive and destructive interference. When the phase difference between the two modes is 180° ($\Delta \varphi = \pi$) all the power is transferred to the other waveguide. Therefore, along the coupled region, the mode can perform a number of jumps equal to $N$ with an overall phase accumulation of $\pi N$. Thus, Eq.~(\ref{eq:powersplit}) can be rewritten as: 
\begin{equation}
\label{eq:powersplit_phase}
    \dfrac{P_{1out}}{P_{1out} + P_{2out}} = cos^2\bigg(\dfrac{\Delta \varphi}{2} \bigg)
\end{equation}
The coupling length $L_c$ depends on the initial dispersion curve and the nonlinear frequency shift coefficient, which in turn depends on the spin-wave power within the directional coupler. If the waveguide considered is not ideal but with losses, the spin-wave power is not constant along the propagation direction. We consider an exponential decay length $e^{(-|2x|/x_{freepath})}$. As a consequence, Eq.~(\ref{eq:powersplit}) is not sufficient because it only considers a constant coupling length, while $L_c$ is continuously varying along the waveguide due to the space-dependent spin-wave power. The equation is rewritten introducing the concept of average coupling length ($L_{a,avg}$) according to Eq.~(\ref{eq:lc_avg}),
\begin{equation}
\label{eq:lc_avg}
    L_{a,avg} = \dfrac{\pi L_w}{\Delta \varphi}
\end{equation}

The phase accumulated between the two modes can be obtained by integrating the wavenumber variation along the propagation direction $\Delta \varphi = \int |k_s - k_{as}| dx$. In our model, the directional coupler is discretized along $x$ direction with step size of $\Delta x$ in a total number of $M$. As a consequence, the phase accumulated $\Delta \varphi$ can be computed as:
\begin{equation}
\label{eq:phase_accumulated}
    \Delta \varphi = \sum^{M}_{i=1} \Delta k_i \Delta x
\end{equation}
For each subinterval $i$, the difference between the two wavenumbers is recalculated. Additionally, the model considers the coupling introduced by the opening/closing arms. Thus, regions 1 and 3 are discretized, but here the gap is varying. For those regions, the calculation starts from the unshifted dispersion relation, which depends on the gap $\sigma$. In this way, the cumulative phase accumulation makes it possible to correctly estimate the average coupling length and, therefore, the output power partition.
\subsection*{Magnonic full adder design with regenerators}
\begin{figure}[ht]
\centering
  \includegraphics[width=\linewidth]{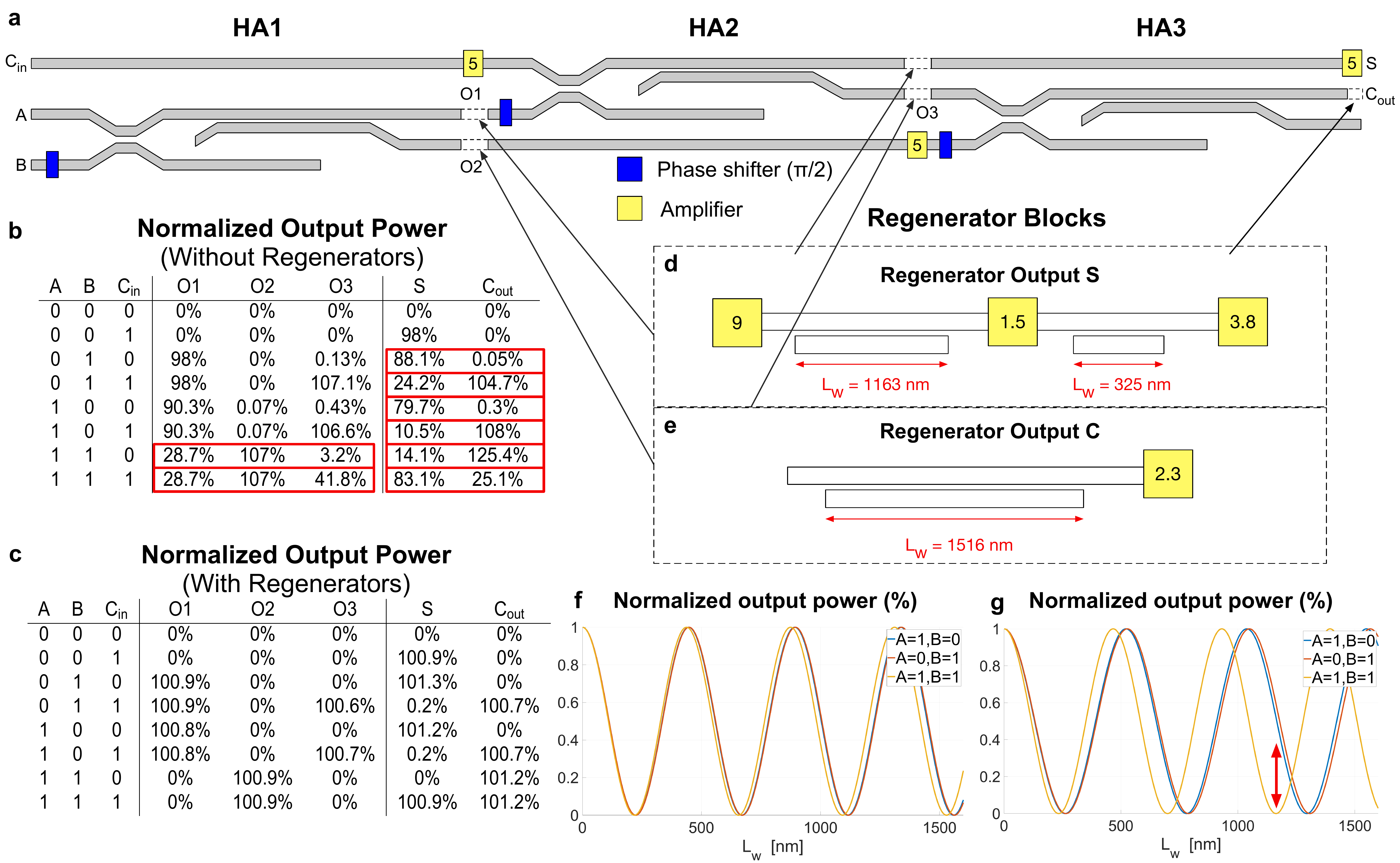}
  \caption{The layout of the design FA and regenerators: \textbf{a}) Full adder composed of three HA. Blue squares identify the phase-shifter block, while the yellow ones are the amplifiers. The outputs of the single HA are identified with labels; \textbf{b}) and \textbf{c}) Tables with the power distributions at the outputs of the FA and HAs without and with the regenerators. The values equal to 0\% refer to the normalized output power lower than 1e-3\%. \textbf{d}) and \textbf{e}) Layout and physical dimensions of the designed regenerator blocks with the 100\,nm YIG node. The gap $\sigma$ in all the  regenerators is 10\,nm; \textbf{f}) and \textbf{g}) Normalized output power as a function of the coupling length of DC2 of an HA for all the input combinations. \textbf{f} shows that it is not possible to find a coupling length that separate the case $A=B=1$ ($S=0$) from the other ($S=1$); \textbf{g}) Presents the same plot after amplification by a factor $9$ that increases the separation between the two logic values.}
  \label{fig:fig3} 
\end{figure}

The adoption of the aforementioned simulation model enabled the design of more complex structures. Cascading the HA\supercite{Wang2020} introduces degradation of the signals, which results in errors in the logical evaluation. In magnonic circuits, the power available at the output identifies the logic values. In particular, a signal below $\nicefrac{1}{3}$ of the \textit{Logic 1} power is considered as \textit{Logic 0}. Amplifiers are used to restore the signals. However, after the amplification, some correctly evaluated values are fed to the following block, causing errors. The usage of amplifiers is not enough to guarantee correct information propagation. Figure \ref{fig:fig3}.b shows the normalized power available at different internal nodes of the FA depicted in Figure \ref{fig:fig3}.a. The values highlighted in red show outputs that can cause errors if directly applied to the following blocks. These errors are because the HA outputs, when amplified and put as input to the following stage reduced the separation among logic values given the extreme non-linearity of DC2. To overcome this limitation, an additional element was introduced in the circuit, the regenerators. Thanks to the non-linearity of specifically designed directional couplers, these blocks increased the logical separation of the outputs. Figure \ref{fig:fig3}.c shows the output of the same circuit when the regenerators are considered. 

The main functionality of the regenerators is to increase the guarantee correct logic values exploiting the magnonic characteristic. As presented in the previous section, the directional couplers have a strong non-linearity, and they can be used to improve signal integrity. The idea is to design the length of the coupler to reduce the amplitude of the \textit{Logic 0} outputs before the amplifier stage. By selecting the correct physical length, it is possible to attenuate the values close to "0" and amplify the "1". In this way, when the signal is amplified, the "1" is correctly restored to $100\%$ energy, while the "0" is still close to $0\%$ energy. The new DC is similar to DC2, operating in the non-linear regime, but only one output branch is considered. The other is used to dissipate unwanted power.

The design methodology is presented here using the HA as a case study. The output power available at the output \textit{S} of the HA are named based on the input combination (case A = '0' and B = '0' is not reported since it refers to no input power):
\begin{itemize}{}
    \item $S10$: when the HA input combination is A = '1' and B = '0'.
    \item $S01$: when the HA input combination is A = '0' and B = '1'.
    \item $S11$: when the HA input combination is A = '1' and B = '1'.
\end{itemize}
The normalized powers before the amplification are $S10$ = $21.5\%$, $S01$ = $23.3\%$ and $S11$ = $6.8\%$. Figure \ref{fig:fig3}.f shows the curves for each output power over the length of the DC. The idea is to find the best length to ensure that the regenerator input power (the power of the output S) is entirely conserved for $S10$ and $S01$, but it is completely transferred to the dummy branch of the coupler for $S11$. Since the curves represent the percentage of power at the useful output of the DC, the best point on the curves is where $S10$ is almost 0, and the other powers are maximum. Unfortunately, the three curves are almost completely superposed, and it was not possible to find a suitable length for the new DC implementing the regenerator. An amplifier of value nine was inserted to increase the separation of the curves. In this way, the spin-wave powers increase up to $S10$ = $193.5\%$, $S01$ = $209.9\%$ and $S11$ = $61.56\%$. With the new values, it was possible to select a length of $1163$\,nm that resulted in the complete attenuation of $S11$ as showed in Figure \ref{fig:fig3}. However, the drawback of the inserted DC was that $S10$ and $S01$ became  $85\%$ and $105\%$ respectively. An additional DC was inserted to mitigate this difference: an amplifier by factor 1.5, followed by a DC $325$\,nm long and a final 3.8x amplifier. This final structure resulted in very similar output power for both signals $S10$ and $S01$, around $100\%$, and zero power for signal $S11$.

The same approach was used to design the regenerator for output C: a single DC with length $1516$\,nm and an amplifier by factor 2.3 was inserted in this case.

Finally, an amplifier was inserted after a long piece of waveguide used to interconnect, for example, \textit{O2} with the input of the last HA. The attenuation of the signal was evaluated using the model and compensated to ensure a correct evaluation of the outputs. Furthermore, a phase shift of \nicefrac{$\pi$}{2} is needed at one input of each HA\supercite{Wang2020}. A specifically designed block, which introduces a geometrical restriction of the waveguide \supercite{Dobrovolskiy2019,Au2012}, was placed in the design to ensure the phase shift.
This design shows the need to restore logic 0 and logic 1 to the proper value to ensure the correct signal propagation along the circuit. The propagation of non-restored signals may result in errors in the computation after few elaboration phases.

%

\subsection*{Scaling and performance analysis}
\begin{figure}[ht]
\centering
  \includegraphics[width=\linewidth]{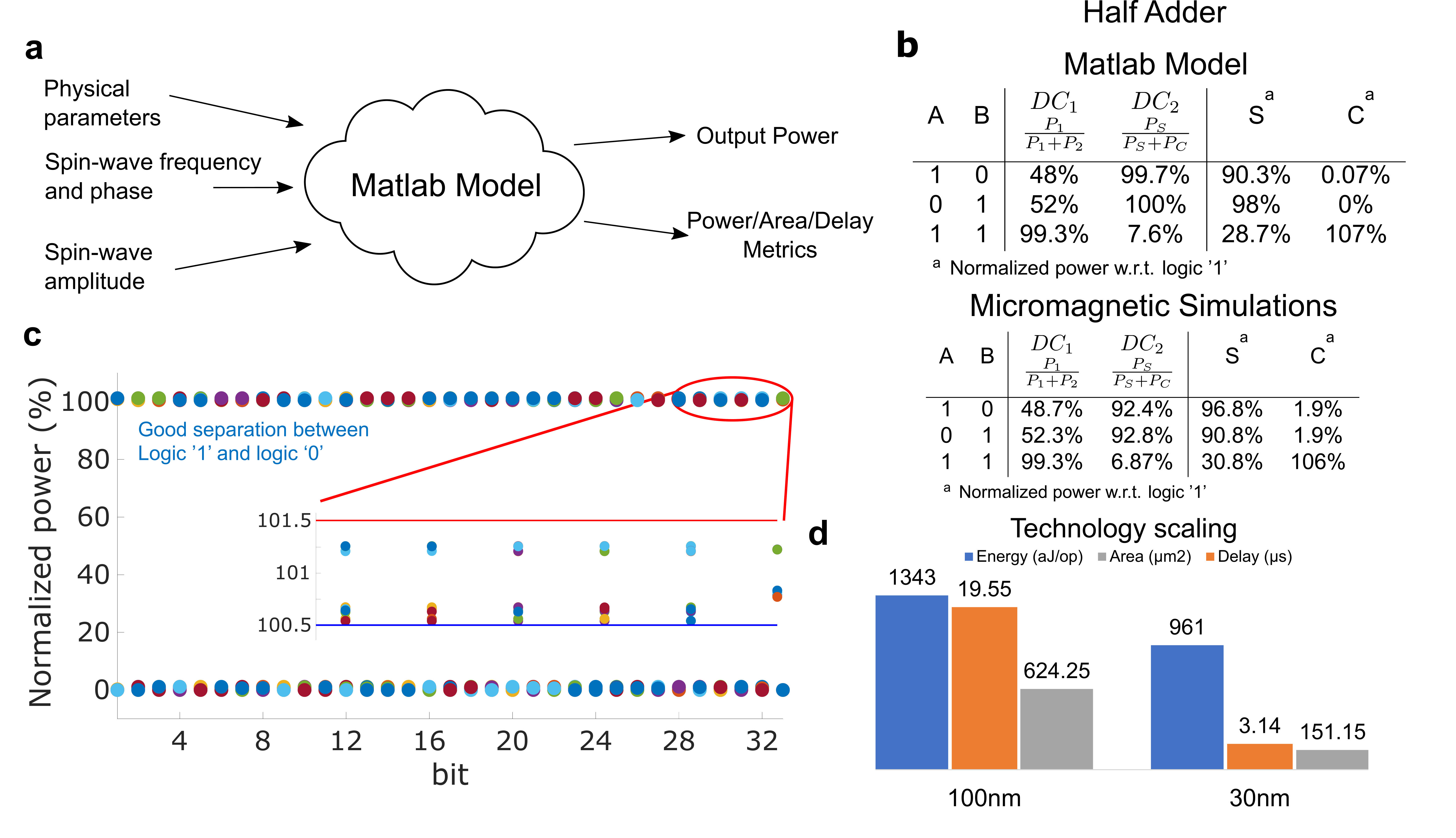}
  \caption{Magnonic circuits metrics and results thanks to the presented model; \textbf{a}) Overview of the model showing required inputs and produced outputs; \textbf{b}) Comparison among the output power distribution obtained with micromagnetic simulations and the proposed Matlab model; \textbf{c}) The output power distribution of a 32-bit RCA repeated 50 times with random inputs. The zoomed portion highlights that thanks to the regenerators, the output evaluated as \textit{Logic 1} are very close to the $100\%$ of the power; \textbf{d} Metrics obtained with the presented model for two technology nodes: 100\si{\nm} and 30\si{\nm}.}
  \label{fig:fig4} 
\end{figure}

The model presented was developed considering two technology nodes, the 100\,nm and the 30\,nm, where the node represents the waveguide width. 
Figure \ref{fig:fig4}.a shows the general idea of the developed MatLab model. It takes as input the material parameters and the geometry of the waveguides, the spin-wave frequency, phase, and amplitude. Different circuit topologies were already defined inside the model code. After the computation, the normalized output power for every input and area/delay/power metrics are reported. The former is used to evaluate the correct behavior of the selected circuit, while the latter can be used to evaluate the performance of the technology. 
Figure \ref{fig:fig4}.b compares the results of the simulations of the HA using the presented MatLab model and the micromagnetic simulations. 
Both models result in similar power distributions at the outputs of the DCs. Columns $S$ and $C$ show the normalized output power with respect to the power of the \textit{Logic 1}. The Matlab simulations were performed considering the regenerators, therefore, the output powers are slightly different with respect to the micromagnetic ones.

A more complex circuit could not be simulated with the micromagnetic simulator due to the huge computational cost and storage limitation in the computer. Our model provides a simple solution by solving the analytical theory with appropriate approximations. Figure \ref{fig:fig4}.c shows the output power distribution of a 32-bit ripple carry adder(RCA). In the picture, 50 calculations with random inputs are depicted. Each dot represents the output power, normalized with respect to \textit{Logic 1}, of each stage of the RCA (32 sum bit and the final carry out). It can be noticed that a perfect separation between "0" and "1" is obtained. Furthermore, the zoomed graph highlights that all the outputs evaluated as \textit{Logic 1} are in the $100\pm1.5\%$ range thanks to the introduction and modeling of the regenerator blocks.

Two different magnonic technology nodes were developed and inserted in the model: 100\si{\nm} and 30\si{\nm} waveguides\supercite{Wang2020}. The model easily gives the possibility to compare the two technology nodes under various aspects. Figure \ref{fig:fig4}.d shows the energy consumption, the area and the delay metrics for the 32-bit RCA in the two technology nodes. It is possible to notice that the scaling from 100\si{\nm} to 30\si{\nano \meter} YIG technology resulted in various benefits. Energy passed from 1343 \si{\atto \joule \per op} to 961 \si{\atto \joule \per op}, meaning that the circuit is more power-efficient. Similarly, area occupation for the RCA dropped from 624.25 \si{\micro \meter \squared} in the 100\si{\nm} to 151.15 \si{\micro \meter \squared} in YIG 30\si{\nm}, resulting in an $76\%$ improvement. Here, the delay is the time needed for the spin-wave to propagate from input to output. It resulted in 19.55 \si{\micro \second} for the YIG 100\si{\nm} and 3.14 \si{\micro \second} for the scaled node. 

The presented model allowed for the evaluation of the performance improvement of technological scaling.  Furthermore, the design of a complex circuit showed the necessity of restoring the logic signals.
%

\subsection*{Methods}
\label{sec:method}
\subsubsection*{Dispersion relation}
The numerical model developed considers two technology nodes, the 100\,nm and the 30\,nm, where the node represents the waveguide width. We considered the following parameters for YIG: the saturation magnetization $M_s=\SI{1.4e5}{\ampere\per\meter}$, the damping $\alpha = \num{2e-4}$ and exchange stiffness $A = \SI{3.5e-12} {\joule \per \meter}$.
The dispersion relation of the spin-wave mode in an isolated waveguide is expressed by Eq.~(\ref{eq:f_0}) according to the work in \cite{Wang2018}.
\begin{equation} \label{eq:f_0}
\begin{split}
    f_0(k_x) & = \frac{1}{2\pi}\sqrt{\Omega^{yy} \Omega^{zz}}\\
    & = \frac{1}{2\pi}\sqrt{(\omega_H+\omega_M(\lambda^2k_x^2+F_{k_x}^{yy}(0)))(\omega_H+\omega_M(\lambda^2k_x^2+F_{k_x}^{zz}(0)))}
\end{split}
\end{equation}
where: 
\begin{itemize}{}
   \item $\Omega^{ii}$ = $ \omega_H+\omega_M(\lambda^2k_x^2+F_{k_x}^{ii}(0))$, $i$=$y,z$.
   \item $\omega_H$ = $\gamma B_{ext}$, $\gamma$ is the gyromagnetic ratio, and $B_{ext}$ is external magnetic field.
   \item $\omega_M$ = $\gamma$ $M_s$, $M_s$ is the saturation magnetization.
   \item $\lambda$ = $\sqrt{2A/(\mu_0 M_s^2)}$ is the exchange length, $A$ is the exchange stiffness, and $\mu_0$ is permeability of vacuum.
   \item $\hat{F}_{k_x}$ is a tensor that describes the dynamical magneto-dipolar interaction. 
\end{itemize}   
The tensor $\hat{F}_{k_x}$ calculation developed by Beleggia et al. can be calculated using the Fourier-space approach\supercite{Beleggia2004}:
\begin{equation} \label{eq:F_d}
      \hat{F}_{k_x}(d) = \frac{1}{2\pi} \int_{-l}^{+l} \hat{N_k}e^{ik_yd} \,dk_y 
\end{equation}

\begin{equation}
       \hat{N_k} = \frac{\left| \sigma_k \right|^2}{\widetilde{w}}
       \begin{pmatrix}
        \frac{k_x^2}{k^2}f(kh) & \frac{k_xk_y}{k^2}f(kh) & 0 \\
        \frac{k_xk_y}{k^2}f(kh) & \frac{k_y^2}{k^2}f(kh) & 0 \\
        0 & 0 & 1-f(kh)
        \end{pmatrix}
\end{equation}
where:
\begin{equation}
        \sigma_k = 2 \frac{k_y cos\Big(\frac{\kappa w}{2}\Big)sin\Big(\frac{k_y w}{2}\Big)-\kappa cos\Big(\frac{k_y w}{2}\Big)sin\Big(\frac{\kappa w}{2}\Big)}{k_y^2-\kappa^2}
\end{equation}
\begin{equation}
       \widetilde{w} = \frac{w}{2}(1+sinc(\kappa w))
\end{equation}
\begin{equation}
       f(kh) = 1-\frac{1-e^{-kh}}{kh}
\end{equation}
\begin{equation}
       k = \sqrt{k_x^2+k_y^2}
\end{equation}
and $h$ is the waveguide thickness, which is equal to \SI{30}{\nm} and \SI{10}{\nm} for the \SI{100}{\nm} and \SI{30}{\nm} technology node respectively. The tensor $\hat{F}_{k_x}(d)$ represents the self-dipolar interation when $d=0$ and represents the dipolar interation between waveguides when computed at distance $d$.
In the case of the isolated waveguide, the $d = 0$ and the integral limitation $l = 10$. Note that in the ideal case, $l$ could be infinite. However, the main contributions of this integral are around $l = 0$.\\
Starting from these considerations, it is possible to obtain the dispersion relation of two coupled waveguides \supercite{Wang2018}. The split between the symmetric and the antisymmetric mode depends on the dipolar interaction and can be computed according to Eq.~(\ref{eq:f_s_as}).
\begin{equation} \label{eq:f_s_as}
    f_{s,as}(k_x) = \frac{1}{2\pi}\sqrt{(\Omega^{yy}\pm \omega_MF_{kx}^{yy}(d)) (\Omega^{zz}\pm \omega_MF_{kx}^{zz}(d))}
\end{equation}
where:
\begin{itemize}{}
   \item $\Omega^{ii}$ = $ \omega_H+\omega_M(\lambda^2k_x^2+F_{k_x}^{ii}(0))$, $i$=$y,z$.
   \item $d$ = $w+\delta$, $w$ is the width of the waveguides, and $\delta$ represents the gap between the two waveguides center to center.
   \item $F_{k_x}^{ii}(d)$ is calculated according to Eq.~(\ref{eq:F_d}).
\end{itemize}
In the coupled waveguides, the profile of the spin-wave is slightly different compared to the single waveguide\supercite{Wang2020}. However, the Eq.~(\ref{eq:f_s_as}) could not take into account the difference. To compensate this error, a solution is to reduce the integral limitation $l$ of Eq.~(\ref{eq:F_d}).\\
The dispersion relation calculation for isolated and coupled waveguides makes it possible to obtain the associated wave number and compute the signal propagating in large circuits.

\subsubsection*{Geometry}
The physical geometries of the diretional couplers depicted in Figure~\ref{fig:fig_1b}.b are strictly related to the adopted technology node. For the 100\,nm YIG the main geometrical quantities for the DC1 are: $L_{w1}$ = 370\,nm, $d_1$ = 450\,nm, $\varphi_1$ = 20\si{\degree}, $\sigma_1$ = 50\,nm. The DC2 is based on the following dimensions: $L_{w2}$ = 3\si{\um}, $d_2$ = 210\,nm, $\varphi_2$ = 20\si{\degree}, $\sigma_2$ = 10\,nm. The sizes involved in the DC1 and DC2 when considering the 30\,nm YIG are: $L_{w1}$ = 230\,nm, $d_1$ = 50\,nm, $\varphi_1$ = 20\si{\degree}, $\sigma_1$ = 20\,nm, $L_{w2}$ = 2460\si{\um}, $d_2$ = 70\,nm, $\varphi_2$ = 20\si{\degree}, $\sigma_2$ = 10\,nm.

\subsubsection*{Metrics}
The physical compact model described to evaluate the signal propagation on every node of the circuit can also be used to extract metrics for analyzing the circuit performance. The model considers the physical geometry of directional couplers, such as the waveguide width and material properties like the gyromagnetic ratio, the damping, the saturation magnetization and the exchange stiffness. The model enables the estimation of the following metrics: occupied area, propagation delay and the energy consumption.\\
The area occupation can be estimated considering the bounding box that encloses every directional coupler.
\begin{equation}
    A_{DC} = w_{DC} * L_{DC}
\end{equation} 
where $w_{DC}$ represents the width of the DC and it is equal to $2w + 4 \cdot 5h$ ($w$ and $h$ are the width and the thickness respectively)\supercite{Wang2020}. The quantity $5h$ is related to the physical geometry and it is used to compute the minimum distance between two waveguides to have negligible dipolar coupling. The quantity $L_{DC}$ refers to the physical length of coupler and can be computed as $L_w + 2 \frac{5h}{sin \varphi}$, where $\varphi$ is the angle of the opening/closing arms of the waveguide.\\
The input-output delay accumulated by every magnonic element can be estimated considering the entire length of every magnonic block divided by the spin-wave group velocity. Being the group velocity dependent on the wave number, the model considers the contribution introduced by the three regions discussed in section~\nameref{sec:model}. Regions 1 and 3 model the propagation delay as dependent by the spin-wave propagating within an isolated waveguide ($k_0$). On the other end, region 2 considers the propagation of the two modes that have a different delay. In general, the contribution from every zone can be computed as $\tau_{zone_i} = \dfrac{L_{zone_i}}{v_{gr_i}}$. To evaluate the overall computation time of a single device, all the contributions are summed together considering the largest delay introduced by region 2:
\begin{equation}
    \tau_{DC} = \tau_{zone_1} + max\{\tau_{zone_2}^s, \tau_{zone_2}^{as}\} + \tau_{zone_3}
\end{equation}
This approach is applied to every directional coupler. For example the HA described in this paper is composed of two directional coupler (DC1 and DC2) and two regenerator blocks, one for the output S and one for the output C. The two regenerators do not have the same length resulting in two different delays:
\begin{align}
    \tau_{HA_S} &= \tau_{DC_1} + \tau_{DC_2} + \tau_{reg_S}\\
    \tau_{HA_C} &= \tau_{DC_1} + \tau_{DC_2} + \tau_{reg_C}
\end{align}
The computed delay is then transferred to the subsequent computing elements up to the output.\\
The energy consumption is calculated as the sum of the spin-wave excitation ($E_{SW}$) and the VCMA amplifier ($E_{amp}$). The energy required to excite the spin-wave and required by the VCMA amplifier was estimated in\supercite{Wang2020} as \SI{12.3}{\atto \joule} and \SI{3}{\atto \joule} per operation, respectively. These quantities refer to the \SI{100}{\nano \meter} node. Scaling the technology to the \SI{30}{\nano \meter} the energy required to excite the spin-wave is reduced to \SI{1.96}{\atto \joule}. When a logic '0' is present at the circuit input, no spin-wave is excited, and no power is dissipated. Therefore, the power consumption depends on the probability of the input to assume a logic '1' (excited spin-wave). For the HA we considered for simplicity an input probability of $P(A='1')=P(B='1')=0.5$. The energy consumption can be computed as:
\begin{equation}
\label{eq:energy_ha}
    E_{HA} = P(A='1') \cdot E_{SW_A} + P(B='1') \cdot  E_{SW_B} + \sum_j^S E_{amp_j}^{HA}
\end{equation}
where $E_{SW_A}, E_{SW_B}$ are the energy of excited spin-wave for input A and B respectively. The last summation considers the contribution of the $S$ amplifiers required by the magnonic HA including the regenerators. The HA is basic building block for construction more complex circuit. Therefore, the Eq.~(\ref{eq:energy_ha}) can be easily extended to other magnonic circuits with $N$ inputs and $M$ HA as:
\begin{equation}
\label{eq:energy_generic}
    E_{HA} =  \sum_i^N + P_i (i = '1') \cdot E_{SW_i} + \sum_j^M E_{amp_j}^{HA}
\end{equation}
where $ E_{SW_i}$ is the $ E_{SW}$ of the i-th input with a probability $P_i$ that it assumes a logic '1'.

\paragraph*{Data availability}
The data and the model that support the findings of this study are openly available on Zenodo \supercite{riente_zenodo_2021}.

\paragraph{Acknowledgments}
The authors would like to thank Huqing Zheng for his support in the development of the model. This research has been supported by ERC Starting Grant 678309 MagnonCircuits.

\paragraph{Author contributions}
U.G. derived the compact model and developed the MatLab implementation. U.G and F.R designed the regenerator blocks and developed the ripple carry adder. 
Q.W. and O.V.D. carried out the micromagnetic simulations to validate the model. 
A.V.C. and F.R. conceived the idea and led this project.
U.G. wrote the manuscript with the help of all the coauthors. All authors contributed to the scientific discussion and commented on the manuscript.

\paragraph{Competing interests}
The authors declare no competing interests.

\printbibliography
\end{document}